\documentclass[12pt,english]{article}
\usepackage[T1]{fontenc}
\usepackage[latin9]{inputenc}
\usepackage{subfigure}
\usepackage{graphicx}
\usepackage{amssymb}

\makeatletter

\providecommand{\tabularnewline}{\\}

\usepackage{a4wide}
\usepackage{psfrag}
\makeatletter

\@addtoreset{equation}{section}

\@addtoreset{figure}{section}

\@addtoreset{table}{section}
\makeatother 

\usepackage{babel}
\makeatother

\begin{document}

\title{\begin{flushright}{\normalsize ITP-Budapest Report No. 636}\end{flushright}\vspace{1cm}Boundary
form factors in finite volume}

\author{M. Kormos$^{1}$%
\thanks{E-mail: kormos@general.elte.hu%
} ~and G. Takács$^{2}$%
\thanks{E-mail: takacs@elte.hu%
}\\
\\
$^{1}$\emph{Institute for Theoretical Physics}\\
\emph{Eötvös University, Budapest}\\
\emph{and}\\
$^{2}$\emph{HAS Research Group for Theoretical Physics}\\
\emph{H-1117 Budapest, Pázmány Péter sétány 1/A.}}

\date{7th April 2008}

\maketitle
\begin{abstract}
We describe the volume dependence of matrix elements of local boundary
fields to all orders in inverse powers of the volume. Using the scaling
boundary Lee-Yang model as testing ground, we compare the matrix elements
extracted from boundary truncated conformal space approach to exact
form factors obtained using the bootstrap method. We obtain solid
confirmation for the boundary form factor bootstrap, which is different
from all previously available tests in that it is a non-perturbative
and direct comparison of exact form factors to multi-particle matrix
elements of local operators, computed from the Hamiltonian formulation
of the quantum field theory. 
\end{abstract}

\section{Introduction}

The investigation of integrable boundary quantum field theories started
with the seminal work of Ghoshal and Zamolodchikov \cite{GZ}, who
set up the boundary R-matrix bootstrap, which makes possible the determination
of the reflection matrices and provides complete description of the
theory on the mass shell.

For the calculation of correlation functions, matrix elements of local
operators between asymptotic states have to be computed. In a boundary
quantum field theory there are two types of operators, the bulk and
the boundary operators, where their names indicate their localization
point. The boundary form factor program for calculating the matrix
elements of local boundary operators between asymptotic states was
initiated in \cite{bffprogram}. The validity of form factor solutions
was checked in the case of the boundary scaling Lee-Yang model by
calculating the two-point function using a spectral sum and comparing
it to the prediction of conformal perturbation theory. In \cite{bffcount}
the spectrum of independent form factor solutions in the scaling Lee-Yang
model and the sinh-Gordon model was compared to the boundary operator
content of the ultraviolet boundary conformal field theory and a complete
agreement was found. Further solutions of the boundary form factor
axioms were constructed and their structure was analyzed for the sinh-Gordon
theory at the self-dual point in \cite{ca1}, and for the $A_{2}$
affine Toda field theory in \cite{ca2}.

While the checks performed so far have confirmed the validity of the
boundary form factor bootstrap proposed in \cite{bffprogram}, they
can still be considered incomplete. The spectral sum evaluated in
that paper only receives a very small contribution from form factors
with more than two particles, thus it cannot be considered as a test
of higher form factor functions, and therefore it does not constitute
a stringent verification of the singularity axioms, which form the
basis of the recursive construction of form factors. The solution
counting in \cite{bffcount} does involve form factors up to arbitrary
number; however, the counting procedure uses only some rough features
of the axioms such as power counting of the polynomials involved,
and the structure of the kernels of the recursion relations.

It is therefore desirable to have a direct comparison of form factors
to matrix elements of local operators evaluated directly from the
boundary quantum field theory in a non-perturbative framework. For
periodic boundary conditions, it was shown recently in \cite{fftcsa1,fftcsa2}
that such a comparison can be achieved by extracting the matrix elements
in finite volume, for which one can use the very efficient truncated
conformal space approach. In this paper we give the extension of this
framework to the boundary case, using the boundary scaling Lee-Yang
model as our paradigmatic example.

The paper is structured as follows. In section 2 we review the necessary
ingredients of the boundary form factor bootstrap and give the form
factor solution in the boundary scaling Lee-Yang model for the boundary
operator with the lowest scaling dimension, up to $6$ particles.
In section 3 we derive a framework which makes it possible to calculate
finite volume matrix elements using the infinite volume form factors,
to all orders in inverse powers of the volume $L$, i.e. neglecting
only corrections that decay exponentially fast with $L$. In section
4 we give the necessary background on the boundary truncated conformal
space method, and also discuss the numerical precision and sources
of errors. The actual comparison between the predictions from the
bootstrap solution and the numerically evaluated finite volume matrix
elements is performed in section 5, and the conclusions are drawn
in section 6.

\section{Boundary form factor bootstrap for the scaling Lee-Yang model}

\subsection{The boundary form factor axioms}

The axioms satisfied by the form factors of a local boundary operator
were derived in \cite{bffprogram}. Here we only list them without
much further explanation. Let us suppose that we treat an integrable
boundary quantum field theory in the (infinite volume) domain $x<0$,
with a single scalar particle of mass $m$, which has a two-particle
$S$ matrix $S(\theta)$ (using the standard rapidity parametrization)
and a one-particle reflection factor $R(\theta)$ off the boundary,
satisfying the boundary reflection factor bootstrap conditions of
Ghoshal and Zamolodchikov \cite{GZ}. For a local operator $\mathcal{O}(t)$
localized at the boundary (located at $x=0$, and parametrized by
the time coordinate $t$) the form factors are defined as\begin{eqnarray*}
\,_{out}\langle\theta_{1}^{'},\theta_{2}^{'},\dots,\theta_{m}^{'}\vert\mathcal{O}(t)\vert\theta_{1},\theta_{2},\dots,\theta_{n}\rangle_{in} & =\\
 &  & \hspace{-2cm}F_{mn}^{\mathcal{O}}(\theta_{1}^{'},\theta_{2}^{'},\dots,\theta_{m}^{'};\theta_{1},\theta_{2},\dots,\theta_{n})e^{-imt(\sum\cosh\theta_{i}-\sum\cosh\theta_{j}^{'})}\end{eqnarray*}
for $\theta_{1}>\theta_{2}>\dots>\theta_{n}>0$ and $\theta_{1}^{'}<\theta_{2}^{'}<\dots<\theta_{m}^{'}<0$,
using the asymptotic $in/out$ state formalism introduced in \cite{BBT}.
They can be extended analytically to complex values of the rapidity
variables. With the help of the crossing relations derived in \cite{bffprogram}
all form factors can be expressed in terms of the elementary form
factors\begin{equation}
\,_{out}\langle0\vert\mathcal{O}(0)\vert\theta_{1},\theta_{2},\dots,\theta_{n}\rangle_{in}=F_{n}^{\mathcal{O}}(\theta_{1},\theta_{2},\dots,\theta_{n})\label{eq:elementaryff}\end{equation}
which can be shown to satisfy the following axioms%
\footnote{There is a further axiom corresponding to boundary excited state poles,
but it will not be needed in the sequel.%
}: 

I. Permutation:

\begin{center}
\begin{equation}
F_{n}^{\mathcal{O}}(\theta_{1},\dots,\theta_{i},\theta_{i+1},\dots,\theta_{n})=S(\theta_{i}-\theta_{i+1})F_{n}^{\mathcal{O}}(\theta_{1},\dots,\theta_{i+1},\theta_{i},\dots,\theta_{n})\label{eq:permutation}\end{equation}

\par\end{center}

II. Reflection:\begin{equation}
F_{n}^{\mathcal{O}}(\theta_{1},\dots,\theta_{n-1},\theta_{n})=R(\theta_{n})F_{n}^{\mathcal{O}}(\theta_{1},\dots,\theta_{n-1},-\theta_{n})\label{eq:reflection}\end{equation}

III. Crossing reflection: \begin{equation}
F_{n}^{\mathcal{O}}(\theta_{1},\theta_{2},\dots,\theta_{n})=R(i\pi-\theta_{1})F_{n}^{\mathcal{O}}(2i\pi-\theta_{1},\theta_{2},\dots,\theta_{n})\label{eq:crossing_reflection}\end{equation}

IV. Kinematical singularity\begin{equation}
-i\mathop{\textrm{Res}}_{\theta=\theta^{'}}F_{n+2}^{\mathcal{O}}(\theta+i\pi,\theta^{'},\theta_{1},\dots,\theta_{n})=\left(1-\prod_{i=1}^{n}S(\theta-\theta_{i})S(\theta+\theta_{i})\right)F_{n}^{\mathcal{O}}(\theta_{1},\dots,\theta_{n})\label{eq:kinematical}\end{equation}

V. Boundary kinematical singularity

\begin{equation}
-i\mathop{\textrm{Res}}_{\theta=0}F_{n+1}^{\mathcal{O}}(\theta+\frac{i\pi}{2},\theta_{1},\dots,\theta_{n})=\frac{g}{2}\Bigl(1-\prod_{i=1}^{n}S\bigl(\frac{i\pi}{2}-\theta_{i}\bigr)\Bigr)F_{n}^{\mathcal{O}}(\theta_{1},\dots,\theta_{n})\label{eq:boundary_kinematical}\end{equation}
where $g$ is the one-particle coupling to the boundary\begin{equation}
R(\theta)\sim\frac{ig^{2}}{2\theta-i\pi}\quad,\quad\theta\sim i\frac{\pi}{2}\label{eq:gdef}\end{equation}

VI. Bulk dynamical singularity \begin{equation}
-i\mathop{\textrm{Res}}_{\theta=\theta^{'}}F_{n+2}^{\mathcal{O}}(\theta+iu,\theta^{'}-iu,\theta_{1},\dots,\theta_{n})=\Gamma F_{n+1}^{\mathcal{O}}(\theta,\theta_{1},\dots,\theta_{n})\label{eq:bulk_dynamical}\end{equation}
corresponding to a bound state pole of the $S$ matrix\[
S(\theta)\sim\frac{i\Gamma^{2}}{\theta-2iu}\quad,\quad\theta\sim2iu\]
(in a theory with a single particle, the only possible value is $u=\pi/3$).

We further assume maximum analyticity i.e. that the form factors have
only the minimal singularity structure consistent with the above axioms.
The general form factor solution can be written in the following form
\cite{bffprogram}\begin{equation}
F_{n}(\theta_{1},\theta_{2},\dots,\theta_{n})=G_{n}(\theta_{1},\theta_{2},\dots,\theta_{n})\prod_{i=1}^{n}r(\theta_{i})\prod_{i<j}f(\theta_{i}-\theta_{j})f(\theta_{i}+\theta_{j})\label{eq:GenAnsatz}\end{equation}
where $f$ is the minimal bulk two-particle form factor satisfying
the conditions\[
f(\theta)=S(\theta)f(-\theta),\qquad f(i\pi+\theta)=f(i\pi-\theta)\]
and having the minimum possible number of singularities in the physical
strip $0\leq\theta<\pi$ together with the slowest possible growth
at infinity \cite{KW}, and $r$ is the minimal boundary one-particle
form factor satisfying\begin{equation}
r(\theta)=R(\theta)r(-\theta)\quad;\quad r(i\pi+\theta)=R(-\theta)r(i\pi-\theta)\label{eq:1pff}\end{equation}
plus analytic conditions similar to those of $f$, but in this case
in the strip $0\leq\theta<\pi/2$. 

The functions $G_{n}$ are totally symmetric and meromorphic in the
rapidities $\theta_{i}$. They are also even and periodic in them
with the period $2\pi i$, so they can only be functions of the variables\[
y_{i}=\mathrm{e}^{\theta_{i}}+\mathrm{e}^{-\theta_{i}}\]
In a theory with only one particle (such as the scaling Lee-Yang model),
the only possible singularity of the $S$ matrix in the physical strip
is located at $\theta=2\pi i/3$ corresponding to the self-fusion
of the particle (plus the crossed channel pole for the same process
at $\pi i/3$) and the relevant fusion coupling is defined as\[
\Gamma^{2}=-i\mathop{\mathrm{Res}}_{\theta=\frac{2\pi}{3}i}S(\theta)\]
Assuming $f$ is chosen such that it has a pole at $\theta=2\pi i/3$
so that it encodes this singularity and that the boundary dynamical
singularities (but not the kinematical ones!) are similarly contained
in the 1PFF function $r$, the functions $G_{n}$ can be written in
the form\[
G_{n}(\theta_{1},\theta_{2},\dots,\theta_{n})=\frac{P_{n}(y_{1},y_{2}\dots,y_{n})}{\prod_{i}y_{i}\,\prod\limits _{i<j}(y_{i}+y_{j})}\]
where the $P_{n}$ are entire functions symmetric in their arguments.
Assuming that the correlation functions involving the operators in
consideration only have power-like short-distance singularities, the
asymptotic growth of the form factors is limited by\[
F_{n}(\theta_{1}+\Lambda,\theta_{2}+\Lambda,\dots,\theta_{n}+\Lambda)\sim\mathrm{e}^{d\Lambda}\]
for some real number $d$, and therefore $P_{n}$ can only be polynomials
of finite degree \cite{bffprogram}.

\subsection{Scaling Lee-Yang model with boundary}

The scaling Lee-Yang model with boundary is a combined bulk and boundary
perturbation of the boundary version of the $\mathcal{M}_{2,5}$ Virasoro
minimal model, which was investigated in detail in \cite{DPTW1}.
The conformal field theory has central charge $c=-22/5$ and the Virasoro
algebra has two irreducible representations $V_{\Delta}$ with highest
weight $\Delta=\Delta_{1,1}=0$ and $\Delta=\Delta_{1,2}=-1/5$. There
is a unique nontrivial relevant bulk perturbation given by the spinless
field $\phi$ with scaling dimensions $\Delta=\bar{\Delta}=-1/5$
which for an appropriate choice of the sign of the coupling flows
to a massive infrared fixed point.

Boundary conformal field theory was developed in \cite{cardybcft,bulkboundaryope,sewing}
and the interested reader is referred to them for details. Applying
the formalism to the conformal Lee-Yang model it can be seen that
there are two conformally invariant boundary conditions. On one of
them, denoted \textbf{$\mathbf{1}$} (i.e. identity) in \cite{DPTW1},
there is no possibility for a boundary perturbation. In the other
case, denoted $\Phi$ in \cite{DPTW1}, there is a nontrivial relevant
boundary field $\varphi$ with scaling dimension $-1/5$ and the general
perturbed boundary conformal field theory action can be written as\begin{equation}
{\mathcal{A}}_{\lambda,\Phi(h)}={\mathcal{A}}_{\Phi}+\lambda\int\limits _{-\infty}^{\infty}dy\int\limits _{-\infty}^{0}dx\phi(x,y)+h\int\limits _{-\infty}^{\infty}dy\varphi(y)\label{eq:slyaction}\end{equation}
where ${\mathcal{A}}_{\Phi}$ denotes the action for ${\mathcal{M}}(2/5)$
with the $\Phi$ boundary condition imposed at $x=0$, and $\lambda$
and $h$ denote the bulk and boundary couplings respectively. The
action of ${\mathcal{A}}_{\lambda,1}$ is similar, but the last term
on the right hand side is missing. For $\lambda>0$ the bulk behaviour
is described by an integrable massive theory having only a single
particle with mass $m$ with the following S matrix \cite{CM}:\[
S(\theta)=-\left(\frac{1}{3}\right)\left(\frac{2}{3}\right)=-\left[\frac{1}{3}\right]\quad;\quad(x)=\frac{\sinh\left(\frac{\theta}{2}+\frac{i\pi x}{2}\right)}{\sinh\left(\frac{\theta}{2}-\frac{i\pi x}{2}\right)}\quad,\quad[x]=(x)(1-x)\]
The minimal bulk two-particle form factor only has a zero at $\theta=0$
and a pole at $\theta=\frac{2\pi i}{3}$ in the strip $0\leq\Im m(\theta)<\pi$
and is of the form \cite{Z1}:\begin{equation}
f(\theta)=\frac{y-2}{y+1}v(i\pi-\theta)v(-i\pi+\theta)\quad,\quad y=e^{\theta}+e^{-\theta}\label{eq:min2pffly}\end{equation}
 where\[
v(\theta)=\exp\left\{ 2\int_{0}^{\infty}\frac{dt}{t}e^{i\frac{\theta t}{\pi}}\frac{\sinh\frac{t}{2}\sinh\frac{t}{3}\sinh\frac{t}{6}}{\sinh^{2}t}\right\} \]
For the perturbed $\Phi$ boundary, the reflection amplitude of the
particle depends on the boundary coupling constant as \cite{DPTW1}\[
R(\theta)_{\Phi}=\left(\frac{1}{2}\right)\left(\frac{1}{6}\right)\left(-\frac{2}{3}\right)\left[\frac{b+1}{6}\right]\left[\frac{b-1}{6}\right]\]
where the dimensionless parameter $b$ is related to the dimensionful
$h$ as \cite{dtwgfunct} \begin{eqnarray}
h(b) & = & -h_{crit}\, m^{6/5}\sin\left[\left(b+\frac{1}{2}\right)\frac{\pi}{5}\right]\label{eq:hbrelation}\\
 &  & h_{crit}=\pi^{\frac{3}{5}}2^{\frac{4}{5}}5^{\frac{1}{4}}\frac{\sin\frac{2\pi}{5}}{\sqrt{\Gamma(\frac{3}{5})\Gamma(\frac{4}{5})}}\left(\frac{\Gamma(\frac{2}{3})}{\Gamma(\frac{1}{6})}\right)^{\frac{6}{5}}=0.685289983991\dots\nonumber \end{eqnarray}
and $m$ is the mass of the particle giving the overall scale in the
infrared description, which is related to the bulk coupling $\lambda$
as \cite{lytba,massscale}\begin{eqnarray}
m & = & \kappa\lambda^{5/12}\label{eq:massgap}\\
 &  & \kappa=2^{19/12}\sqrt{\pi}\frac{\left(\Gamma(3/5)\Gamma(4/5)\right)^{5/12}}{5^{5/16}\Gamma(2/3)\Gamma(5/6)}=2.64294466304\dots\nonumber \end{eqnarray}
In the case of the $\mathbf{1}$ boundary the reflection amplitude
is the parameter independent expression\[
R(\theta)_{1}=\left(\frac{1}{2}\right)\left(\frac{1}{6}\right)\left(-\frac{2}{3}\right)\]

\subsection{Recursion relations and form factor solution for the $\Phi$ boundary
condition}

The minimal boundary one-particle form factor is \cite{bffprogram}\begin{equation}
r_{\Phi}(\theta)=\frac{i\sinh\theta}{(\sinh\theta-i\sin\frac{\pi(b+1)}{6})(\sinh\theta-i\sin\frac{\pi(b-1)}{6})}\,\, u(\theta)\label{eq:1pffphi}\end{equation}
where\[
u(\theta)=\exp\left\{ \int_{0}^{\infty}\frac{dt}{t}\left[\frac{1}{\sinh\frac{t}{2}}-2\cosh\frac{t}{2}\cos\left[\left(\frac{i\pi}{2}-\theta\right)\frac{t}{\pi}\right]\frac{\sinh\frac{5t}{6}+\sinh\frac{t}{2}-\sinh\frac{t}{3}}{\sinh^{2}t}\right]\right\} \]
Taking the following Ansatz\begin{equation}
F_{n}(\theta_{1},\dots,\theta_{n})=N\, H_{n}Q_{n}(y_{1},\dots,y_{n})\prod_{i}\frac{r_{\Phi}(\theta_{i})}{y_{i}}\prod_{i<j}\frac{f(\theta_{i}-\theta_{j})f(\theta_{i}+\theta_{j})}{y_{i}+y_{j}}\label{AnsatzPhi}\end{equation}
where\begin{equation}
H_{n}=\left(\frac{i3^{\frac{1}{4}}}{2^{\frac{1}{2}}v(0)}\right)^{n}\label{eq:Hn}\end{equation}
and $N$ is a normalization constant to be fixed later, the (\ref{eq:bulk_dynamical})
bulk dynamical ($\mathcal{D}$), (\ref{eq:kinematical}) bulk kinematical
($\mathcal{K}$) and (\ref{eq:boundary_kinematical}) boundary kinematical
($\mathcal{B}$) singularity axioms give the following recursion relations
for the polynomials $Q_{n}$ \cite{bffprogram,bffcount}\begin{eqnarray*}
\mathcal{D}: &  & Q_{2}(y_{+},y_{-})=(y^{2}-3+\beta)Q_{1}(y)\\
 &  & Q_{n+2}(y_{+},y_{-},y_{1},\dots,y_{n})=(y^{2}-3+\beta)\, D_{n}(y|y_{1},\dots y_{n})Q_{n+1}(y,y_{1},\dots,y_{n})\quad,\, n>0;\\
\mathcal{K}: &  & Q_{2}(-y,y)=0\\
 &  & Q_{n+2}(-y,y,y_{1},\dots,y_{n})=(y^{4}-(3+\beta)y^{2}+\beta^{2})K_{n}(y|y_{1},\dots y_{n})Q_{n}(y_{1},\dots,y_{n})\quad,\, n>0;\\
\mathcal{B}: &  & Q_{1}(0)=0\\
 &  & Q_{n+1}(0,y_{1},\dots y_{n})=\beta B_{n}(y_{1},\dots y_{n})Q_{n}(y_{1},\dots,y_{n})\quad,\quad n>0;\end{eqnarray*}
where we introduced the parameter\[
\beta=1+2\cos\frac{\pi b}{6}\]
and $D_{n}$, $K_{n}$ and $B_{n}$ are given by \[
K_{n}(y|y_{1},\dots y_{n})=\frac{1}{2(y_{+}-y_{-})}\left[\prod_{i=1}^{n}(y_{i}-y_{-})(y_{i}+y_{+})-\prod_{i=1}^{n}(y_{i}+y_{-})(y_{i}-y_{+})\right]\]
\[
B_{n}(y_{1},\dots,y_{n})=\frac{1}{2\sqrt{3}}\left(\prod_{i=1}^{n}(y_{i}+\sqrt{3})-\prod_{i=1}^{n}(y_{i}-\sqrt{3})\right)\]
\[
D_{n}(y|y_{1},\dots y_{n})=\prod_{i=1}^{n}(y+y_{i})\]
where\begin{eqnarray*}
y_{+} & = & \omega z+\omega^{-1}z^{-1}\\
y_{-} & = & \omega^{-1}z+\omega z^{-1}\qquad,\qquad\omega=\mathrm{e}^{\frac{i\pi}{3}}\end{eqnarray*}
with the auxiliary variable $z$ defined as a solution of $y=z+z^{-1}$
(i.e. writing $y=2\cosh\theta$ we obtain $z=\mathrm{e}^{\theta}$).
The symmetry of the above expressions in $y_{\pm}$ ensures that the
resulting relations only depend on $y$ and also that $K_{n}$ is
a polynomial in all of its variables \cite{bffcount}. 

The polynomials $Q_{n}$ can be expanded using the elementary symmetric
polynomials defined by the generating function\[
\prod_{i=1}^{n}(z+y_{i})=\sum_{k=0}^{n}z^{n-k}\sigma_{k}(y_{1},\dots,y_{n})\]
The solution of the recursion relations $\mathcal{D}$, $\mathcal{K}$
and $\mathcal{B}$ proceeds as follows. All the recursion coefficients
$D_{n}$, $K_{n}$ and $B_{n}$ can be expanded in products of symmetric
polynomials of the variables $y_{1},\dots,y_{n}$ and, for the first
two, powers of the additional variable $y$. For the minimal solution
of these equations the polynomials $Q_{n}$ have degree $n(n+1)/2$
\cite{bffprogram}, so we can write an Ansatz as a general linear
combination of products of elementary symmetric polynomials not exceeding
this degree. Using the relation\[
\sigma_{k}(y,y_{1},\dots,y_{n})=\sigma_{k}(y_{1},\dots,y_{n})+y\sigma_{k-1}(y_{1},\dots,y_{n})\]
we can expand both sides of the recursion relations in terms of products
of $\sigma_{k}$ and powers of $y$. Equating the coefficients of
independent terms on the two sides, we obtain a linear system of equations
for the coefficients of the Ansatz, which can then be solved. The
minimal solution turns out to be unique (apart from an overall normalization
that can be fixed by choosing the coefficient of $Q_{1}$), and up
to $6$ particles it is given by\begin{eqnarray*}
Q_{1}^{\varphi} & = & \sigma_{1}\\
Q_{2}^{\varphi} & = & \sigma_{1}(\beta+\sigma_{2})\\
Q_{3}^{\varphi} & = & \sigma_{1}\left[\sigma_{3}(\sigma_{2}+3)+\beta\sigma_{1}(\sigma_{2}+\beta)\right]\\
Q_{4}^{\varphi} & = & \sigma_{1}(\sigma_{2}+3)\left[(3\sigma_{1}+\sigma_{3})\sigma_{4}+\beta(\sigma_{2}+3)\sigma_{3}+\beta^{2}\sigma_{1}(\sigma_{2}+\beta)\right]\\
Q_{5}^{\varphi} & = & \sigma_{1}((3+\sigma_{2})(3\sigma_{1}+\sigma_{3})-\sigma_{5})\\
 &  & \times\left[\sigma_{4}(3\beta\sigma_{1}+\beta\sigma_{3}+\sigma_{5})+(\beta-3)\beta(\beta^{2}\sigma_{1}-\sigma_{5})+(3+\sigma_{2})(\beta^{3}\sigma_{1}+\beta^{2}\sigma_{3}+3\sigma_{5})\right]\\
Q_{6}^{\varphi} & = & \sigma_{1}\left[-\sigma_{4}\sigma_{5}-3\sigma_{1}\sigma_{6}+(\sigma_{2}+3)(3\sigma_{1}\sigma_{4}+\sigma_{3}\sigma_{4}-3\sigma_{5}-\sigma_{1}\sigma_{6})+3(\sigma_{2}+3)^{2}(3\sigma_{1}+\sigma_{3})\right]\\
 &  & \times\Big\{\beta(\sigma_{2}+3)(\beta^{3}\sigma_{1}+\beta^{2}\sigma_{3}+3\sigma_{5})+(9\sigma_{1}+3\sigma_{3}+\sigma_{5})(3\sigma_{4}+\sigma_{6})\\
 &  & +(\beta-3)\big[81\sigma_{1}-9\sigma_{5}+\sigma_{4}\sigma_{5}+3\sigma_{1}\sigma_{6}\\
 &  & +(\beta+3)(3\sigma_{1}\sigma_{4}+\sigma_{3}\sigma_{4}-\sigma_{1}\sigma_{6}+(\beta-3)(9\sigma_{1}+\beta^{2}\sigma_{1}-\sigma_{5}))\big]\Big\}\end{eqnarray*}
The first three form factors in this sequence were originally computed
in \cite{bffprogram} where it was demonstrated that this solution
corresponds to the relevant boundary field $\varphi$, which gives
the boundary perturbation according to (\ref{eq:slyaction}). The
complete form factors can be written as \cite{bffprogram} \begin{equation}
F_{n}(\theta_{1},\dots,\theta_{n})=\left\langle \varphi\right\rangle H_{n}Q_{n}^{\varphi}(y_{1},\dots,y_{n})\prod_{i}\frac{r_{\Phi}(\theta_{i})}{y_{i}}\prod_{i<j}\frac{f(\theta_{i}-\theta_{j})f(\theta_{i}+\theta_{j})}{y_{i}+y_{j}}\label{PhiFF}\end{equation}
where\begin{equation}
\langle\varphi\rangle=-\frac{5}{6h_{crit}}\frac{\cos(\frac{\pi b}{6})}{\cos(\frac{\pi}{10}(2b+1))}m^{-1/5}\label{eq:exactvev}\end{equation}
 is the exact expectation value of the boundary operator $\varphi$
\cite{DPTW1}.

\section{Boundary form factors in finite volume}

\subsection{Bethe-Yang equations}

Let us consider an integrable boundary quantum field theory with particles
of species $a=1,\dots,N$ and corresponding masses $m_{a}$. As usual
in two-dimensional field theory, we label asymptotic particles with
their rapidities $\theta$, which gives their energy and momentum
as\[
E_{a}\pm p_{a}=m_{a}\mathrm{e}^{\pm\theta_{a}}\]
We assume the bulk scattering is diagonal and is given by the two-particle
$S$ matrices\[
S_{a_{1}a_{2}}\left(\theta_{1}-\theta_{2}\right)=\mbox{e}^{i\delta_{a_{1}a_{2}}\left(\theta_{1}-\theta_{2}\right)}\]
We further assume that the reflection on the boundaries is also diagonal
and is given by the reflection factors \begin{equation}
R_{a}^{(\alpha)}\left(\theta\right)=\mbox{e}^{i\delta_{a}^{(\alpha)}\left(\theta\right)}\qquad,\qquad R_{a}^{(\beta)}\left(\theta\right)=\mbox{e}^{i\delta_{a}^{(\beta)}\left(\theta\right)}\label{eq:bphaseshifts}\end{equation}
where $\alpha$ and $\beta$ denote the left and right boundary conditions,
respectively.

In the diagonal case, the multi-particle energy levels in a finite
volume $L$ are described by the following Bethe-Yang equations \cite{bbye}:\begin{eqnarray}
Q_{j}\left(\theta_{1},\dots,\theta_{n}\right)_{a_{1}\dots a_{n}} & = & 2\pi I_{j}\label{eq:bbye}\end{eqnarray}
where the phases describing the wave function monodromy are given
by\[
Q_{j}\left(\theta_{1},\dots,\theta_{n}\right)_{a_{1}\dots a_{n}}=2m_{a_{j}}L\sinh\theta_{j}+\sum_{k\neq j}\left(\delta_{a_{j}a_{k}}\left(\theta_{j}-\theta_{k}\right)+\delta_{a_{j}a_{k}}\left(\theta_{j}+\theta_{k}\right)\right)+\delta_{a_{j}}^{(\alpha)}\left(\theta_{j}\right)+\delta_{a_{j}}^{(\beta)}\left(\theta_{j}\right)\]
Here all rapidities $\theta_{j}$ (and accordingly all quantum numbers
$I_{j}$) are taken to be positive%
\footnote{Boundary reflections change the sign of the momentum, so finite volume
multi-particle states can be characterized by the absolute value of
the rapidities.%
}. We can take the quantum numbers ordered as $I_{1}\leq\dots\leq I_{n}$;
in the scaling Lee-Yang model they must all be different due to the
exclusion principle. We denote the corresponding multi-particle state
by \[
\vert\{I_{1},\dots,I_{n}\}\rangle_{a_{1}\dots a_{n},L}\]
and its energy (relative to the ground state) is given by\[
E_{I_{1}\dots I_{n}}(L)=\sum_{j=1}^{n}m_{a_{j}}\cosh\tilde{\theta}_{j}\]
where $\left\{ \tilde{\theta}_{j}\right\} _{j=1,\dots,n}$ is the
solution of eqns. (\ref{eq:bbye}) at the given volume $L$. The Bethe-Yang
equations gives the energy of the multi-particle states to all order
in $1/L$, neglecting only finite size effects decaying exponentially
with $L$.

\subsection{Non-diagonal matrix elements}

Following the ideas outlined in \cite{fftcsa1,fftcsa2} we begin with
examining a two-point function of two boundary operators\[
\left\langle \mathcal{O}_{1}(\tau)\mathcal{O}_{2}(0)\right\rangle _{L}\]
where $\tau$ is Euclidean time. Let us suppose for definiteness that
we consider operators located on the right boundary.

We first need to establish that the finite size corrections to the
two-point function will then be of the form\begin{equation}
\left\langle \mathcal{O}_{1}(\tau)\mathcal{O}_{2}(0)\right\rangle _{L}-\left\langle \mathcal{O}_{1}(\tau)\mathcal{O}_{2}(0)\right\rangle \sim\mathrm{e}^{-\mu L}\label{eq:twopt_fincorr}\end{equation}
Unfortunately, for the boundary situation there is no analogue of
Lüscher's systematic finite volume expansion \cite{luscher_onept}
that was used for periodic boundary conditions. Let us recall that
the essential input in the argument is the analytic structure of Green's
function, which was worked out for the boundary case in \cite{BBT};
just as in the bulk case, the singularities are given by diagrams
with all internal lines on-shell. It can be seen from Lüscher's derivation
that the correction results from the singularities of momentum space
Green's functions which occur in the momentum integrals of the finite
volume expansion, and the exponent $\mu$ (which is on the scale of
the mass gap of the theory) is given by the singularity lying closest
to the physical domain. This argument cannot be considered a proper
derivation; it would go much beyond the scope of this paper and requires
developing the finite volume expansion for the boundary case. However,
we still have a strong reason to accept that (\ref{eq:twopt_fincorr})
holds; in retrospect, this is also confirmed by the numerical results
in section 5.

Using (\ref{eq:twopt_fincorr}) we can then follow the arguments outlined
in \cite{fftcsa1} without essential modifications. Just as in the
periodic case we find that the infinite volume and finite volume matrix
elements are just related by the square root of the ratio of normalization
of the corresponding states (up to the corrections neglected in (\ref{eq:twopt_fincorr})).
This results in the following relation:\begin{equation}
\langle0\vert\mathcal{O}(0)\vert\{I_{1},\dots,I_{n}\}\rangle_{a_{1}\dots a_{n},L}=\frac{F_{a_{1}\dots a_{n}}^{\mathcal{O}}(\tilde{\theta}_{1},\dots,\tilde{\theta}_{n})}{\sqrt{\rho_{a_{1}\dots a_{n}}(\tilde{\theta}_{1},\dots,\tilde{\theta}_{n})}}+O(\mathrm{e}^{-\mu L})\label{eq:ffrelation}\end{equation}
where $F_{a_{1}\dots a_{n}}^{\mathcal{O}}(\tilde{\theta}_{1},\dots,\tilde{\theta}_{n})$
is the form factor of the operator $\mathcal{O}$ (in the infinite
volume theory, i.e. on the half-line $x<0$),$\left\{ \tilde{\theta}_{j}\right\} _{j=1,\dots,n}$
is the solution of eqns. (\ref{eq:bbye}) at the given volume $L$,
and \begin{equation}
\rho_{a_{1}\dots a_{n}}(\tilde{\theta}_{1},\dots,\tilde{\theta}_{n})=\det\left\{ \frac{\partial Q_{k}(\theta_{1},\dots,\theta_{n})_{a_{1}\dots a_{n}}}{\partial\theta_{l}}\right\} _{k,l=1,\dots,n}\label{eq:dosfinvol}\end{equation}
is the finite volume density of states, which is the Jacobi determinant
of the mapping between the space of quantum numbers and the space
of rapidities given by the Bethe-Yang equations (\ref{eq:bbye}).
Using the crossing relations derived in \cite{bffprogram} and following
the arguments of \cite{fftcsa1}, general matrix elements can be written
as\begin{eqnarray}
 &  & \,_{b_{1}\dots b_{m}}\langle\{I_{1}',\dots,I_{m}'\}\vert\mathcal{O}(0)\vert\{I_{1},\dots,I_{n}\}\rangle_{a_{1}\dots a_{n},L}=\nonumber \\
 &  & \qquad\frac{F_{\bar{b}_{m}\dots\bar{b}_{1}a_{1}\dots a_{n}}^{\mathcal{O}}(\tilde{\theta}_{m}'+i\pi,\dots,\tilde{\theta}_{1}'+i\pi,\tilde{\theta}_{1},\dots,\tilde{\theta}_{n})}{\sqrt{\rho_{a_{1}\dots a_{n}}(\tilde{\theta}_{1},\dots,\tilde{\theta}_{n})\rho_{b_{1}\dots b_{m}}(\tilde{\theta}_{1}',\dots,\tilde{\theta}_{m}')}}+O(\mathrm{e}^{-\mu L})\label{eq:genffrelation}\end{eqnarray}
as long as the sets of the rapidities corresponding to the two states,
$\left\{ \tilde{\theta}_{j}\right\} _{j=1,\dots,n}$ and $\left\{ \tilde{\theta}_{j}'\right\} _{j=1,\dots,m}$,
are disjoint i.e. when there are no disconnected contributions. Here
$\bar{b}_{j}$ benotes the particle species conjugate to $b_{j}$.

The exponent $\mu$ in (\ref{eq:ffrelation},\ref{eq:genffrelation})
can also be estimated using the Poisson summation formula. The argument
for the periodic case was written down in subsection 2.2 of ref. \cite{fftcsa1}
and carries over without essential modification. The result is that
the deviation between the discrete and continuous versions of the
spectral sums is determined by the analytic singularity structure
of the form factors and the phase-shifts involved in (\ref{eq:ffrelation},\ref{eq:genffrelation})
and the leading correction is given by the one with the smallest imaginary
part in rapidity variables, just as the correction indicated in (\ref{eq:twopt_fincorr}).
Taking the two sources of exponential corrections together the result
is that it is the singularity closest to the physical domain which
determines the leading exponential correction in (\ref{eq:ffrelation},\ref{eq:genffrelation}).
In fact this was already tacitly assumed by using the same exponent
$\mu$ as in (\ref{eq:twopt_fincorr}). This is essentially the same
result that was obtained by Lüscher for mass corrections which periodic
boundary conditions in finite volume \cite{luscher_onept}. For the
case of periodic boundary conditions such corrections to finite volume
scattering states and form factors were recently investigated in \cite{balazs}.

Note that the determinant (\ref{eq:dosfinvol}) has the following
behaviour for large $L$:\[
\rho_{a_{1}\dots a_{n}}(\tilde{\theta}_{1},\dots,\tilde{\theta}_{n})=\left(\prod_{l=1}^{n}2m_{a_{l}}L\,\cosh\tilde{\theta}_{l}\right)\left(1+O\left(L^{-1}\right)\right)\]
The leading terms is just the density of states for non-interacting
particles, and it is independent of the bulk and boundary phase-shifts.
The corrections are of the order of $1/(mL)$ where $m$ is the typical
mass scale. In the numerical examples given in section 5 this means
that the interaction corrections are of relative size $\sim10^{-1}$
in the volume range considered there.

\subsection{Diagonal matrix elements}

When disconnected contributions are present, a more careful analysis
is required. As in \cite{fftcsa2} we must start by finding out the
conditions under which there can be equal rapidities in the two states.
It is easy to see that due to the finite size corrections coming from
the Bethe-Yang equations (\ref{eq:bbye}) and the positivity of all
the quantum numbers $I_{j}$ this can only happen when the quantum
number sets happen to be identical, i.e. for the case of diagonal
matrix elements\[
\,_{a_{1}\dots a_{n}}\langle\{I_{1},\dots,I_{n}\}\vert\mathcal{O}(0)\vert\{I_{1},\dots,I_{n}\}\rangle_{a_{1}\dots a_{n},L}\]
According to (\ref{eq:genffrelation}) for this case we have to consider
\[
F_{\bar{a}_{n}\dots\bar{a}_{1}a_{1}\dots a_{n}}(\theta_{n}+i\pi,...,\theta_{1}+i\pi,\theta_{1},...,\theta_{n})\]
Due to the existence of kinematical poles (\ref{eq:kinematical})
the above expression is not well-defined. Let us consider the regularized
version\[
F_{\bar{a}_{n}\dots\bar{a}_{1}a_{1}\dots a_{n}}(\theta_{n}+i\pi+\epsilon_{n},...,\theta_{1}+i\pi+\epsilon_{1},\theta_{1},...,\theta_{n})\]
Just as for bulk form factors, the singular parts of this expression
drop when taking the limits $\epsilon_{i}\rightarrow0$ simultaneously%
\footnote{This can be proved by elementary manipulations involving the bulk
kinematical singularity axiom (\ref{eq:kinematical}).%
}; however, the end result depends on the direction of the limit, i.e.
on the ratio of the $\epsilon_{i}$ parameters. The terms that are
relevant in the limit can be written in the following general form:
\begin{eqnarray}
F_{\bar{a}_{n}\dots\bar{a}_{1}a_{1}\dots a_{n}}(\theta_{n}+i\pi+\epsilon_{n},...,\theta_{1}+i\pi+\epsilon_{1},\theta_{1},...,\theta_{n})=\label{mostgeneral}\\
\prod_{i=1}^{n}\frac{1}{\epsilon_{i}}\cdot\sum_{i_{1}=1}^{n}...\sum_{i_{n}=1}^{n}\mathcal{A}_{i_{1}...i_{n}}^{a_{1}\dots a_{n}}(\theta_{1},\dots,\theta_{n})\epsilon_{i_{1}}\epsilon_{i_{2}}...\epsilon_{i_{n}}+\dots\nonumber \end{eqnarray}
 where $\mathcal{A}_{i_{1}...i_{n}}^{a_{1}\dots a_{n}}$ is a completely
symmetric tensor of rank $n$ in the indices $i_{1},\dots,i_{n}$,
and the ellipsis denote terms that vanish when taking $\epsilon_{i}\rightarrow0$
simultaneously. This tensor can in principle be calculated using an
appropriate version the graphical methods developed in \cite{fftcsa2},
but we do not go into details here as it is not needed in the sequel
(the interested reader is referred to \cite{ftop}).

The connected matrix element can be defined as the $\epsilon_{i}$
independent part of eqn. (\ref{mostgeneral}), i.e. the part which
does not diverge whenever any of the $\epsilon_{i}$ is taken to zero:
\begin{equation}
F_{a_{1}\dots a_{n}}^{c}(\theta_{1},\theta_{2},...,\theta_{n})=n!\,\mathcal{A}_{1\dots n}^{a_{1}\dots a_{n}}(\theta_{1},\dots,\theta_{n})\label{eq:connected}\end{equation}
where the appearance of the factor $n!$ is simply due to the permutations
of the $\epsilon_{i}$. 

For the case of periodic boundary conditions Saleur proposed a way
of expressing a diagonal matrix element in terms of connected form
factors \cite{saleurfiniteT}, on the basis of earlier ideas by Balog
\cite{balogtba} and the Gaudin determinant formula for the inner
product of Bethe states \cite{qism}. This proposal was tested in
\cite{fftcsa2} for periodic boundary conditions against truncated
conformal space, and a spectacular agreement was found. 

Saleur's argument actually seems to be quite general, and so the appropriate
version of his formula for the boundary case can easily be written,
yielding the following conjecture:\begin{eqnarray}
 &  & \,_{a_{1}\dots a_{n}}\langle\{I_{1}\dots I_{n}\}|\mathcal{O}(0)|\{I_{1}\dots I_{n}\}\rangle_{a_{1}\dots a_{n},L}=\label{eq:diaggenrulesaleur}\\
 &  & \frac{1}{\rho_{a_{1}\dots a_{n}}(\tilde{\theta}_{1},\dots,\tilde{\theta}_{n})}\sum_{A\subset\{1,2,\dots n\}}F_{a(A)}^{c}(\{\tilde{\theta}_{k}\}_{k\in A})\tilde{\rho}_{a_{1}\dots a_{n}}(\tilde{\theta}_{1},\dots,\tilde{\theta}_{n}|A)+O(\mathrm{e}^{-\mu L})\nonumber \end{eqnarray}
The summation runs over all subsets $A$ of $\{1,2,\dots n\}$. For
any such subset, we define the corresponding species index set \[
a(A)=\left\{ a_{k}\right\} _{k\in A}\]
and the appropriate sub-determinant\[
\tilde{\rho}_{a_{1}\dots a_{n}}(\tilde{\theta}_{1},\dots,\tilde{\theta}_{n}|A)=\det\mathcal{J}_{A}^{a_{1}\dots a_{n}}(\tilde{\theta}_{1},\dots,\tilde{\theta}_{n})\]
of the $n\times n$ Bethe-Yang Jacobi matrix \begin{equation}
\mathcal{J}_{a_{1}\dots a_{n}}(\tilde{\theta}_{1},\dots,\tilde{\theta}_{n})_{kl}=\frac{\partial Q_{k}(\theta_{1},\dots,\theta_{n})_{a_{1}\dots a_{n}}}{\partial\theta_{l}}\label{eq:jacsubmat}\end{equation}
obtained by deleting the rows and columns corresponding to the subset
of indices $A$. The determinant of the empty sub-matrix (i.e. when
$A=\{1,2,\dots n\}$) is defined to equal $1$ by convention.

There is a simple way to assign an intuitive meaning to all the terms
in (\ref{eq:diaggenrulesaleur}). One needs to consider all possible
ways to partition the particles into two sets, one of which ($A$)
is connected to the local boundary operator $\mathcal{O}$, while
the particles in the complementary set ($A^{\dagger}$) are just contracted
with each other using the inner product. For the latter contraction
we obtain the inner product \[
\,_{a(A^{\dagger})}\langle\{I_{k}\}_{k\in A^{\dagger}}|\{I_{k}\}_{k\in A^{\dagger}}\rangle_{a(A^{\dagger}),L}\quad,\quad A^{\dagger}=\{1,\dots,n\}\setminus A\]
which is given by the appropriate density of states. As before, this
density of states is given by the Jacobian of the quantum number --
rapidity mapping. However, this must be considered in the presence
of the other particles (those in $A$) which contribute to the quantization
relations for the particles in the set $A^{\dagger}$, and is therefore
given by the corresponding sub-determinant of the Jacobian matrix
(\ref{eq:jacsubmat}). 

Finally we remark that the result (\ref{eq:diaggenrulesaleur}) can
also be expressed with the so-called symmetric evaluation of diagonal
matrix elements instead of the connected one as it was done for the
bulk case discussed in \cite{fftcsa2}; however it turns out that
in the boundary case it does not lead to the same simplifications
as in the bulk (cf. \cite{ftop}) and therefore we omit these details
in the present work.

\section{Numerical determination of matrix elements}

\subsection{The boundary truncated conformal space approach}

Truncated conformal space approach (TCSA) was developed by Yurov and
Zamolodchikov \cite{yurov_zamolodchikov}, who used it to describe
the scaling Lee-Yang model in finite volume with periodic boundary
conditions. The boundary extension of TCSA (dubbed BTCSA) was developed
by G. Watts and collaborators and first used in \cite{DPTW1}. A detailed
description of the method can be found in Runkel's PhD thesis \cite{runkel},
which provides a good starting point for developing a numerical algorithm.
We implemented the computations using the symbolic algebra software
\texttt{Mathematica}.

Here we restrict ourselves to specify our conventions. Following \cite{DPTW1}
we used a basis for the operator algebra in which all the structure
constants are real. To simplify matters we specified the left boundary
condition as the identity one ($\mathbf{1}$) and put the $\Phi$
boundary condition to the right end, which is also the position where
our boundary field $\varphi$ is located. In such a case the Hilbert
space consists of a single $V_{-1/5}$ module whose basis vectors
we denote by $|i\rangle$. We truncated this space at various levels,
the highest truncation taken at level $25$. On this finite dimensional
space we computed the matrix elements of the Hamiltonian\[
H=H_{0}+\lambda\int_{0}^{L}\Phi\left(\tau=0,x\right)dx+h\varphi_{R}(\tau=0)\]
We can measure all quantities in units of the bulk particle mass $m$,
and introduce the dimensionless volume variable $l=mL$. The dimensionless
Hamiltonian matrix can then be written in the form\begin{equation}
h_{ij}=\frac{\pi}{l}\left[\left(\Delta_{i}-\frac{c}{24}\right)\delta_{ij}+\kappa'\left(\frac{l}{\pi}\right)^{12/5}\left(G^{-1}B\right)_{ij}+\chi(b)\left(\frac{l}{\pi}\right)^{6/5}\left(G^{-1}B_{R}\right)_{ij}\right]\label{eq:btcsa_ham}\end{equation}
where $\Delta_{i}$ is the conformal weight of the basis vector $|i\rangle$,
$c=-22/5$ is the central charge, the mass gap constant \[
\kappa'=\kappa^{-12/5}=0.097048456298\dots\]
is calculated from (\ref{eq:massgap}), \[
\chi(b)=h_{crit}\,\sin\left(\bigl(b+\frac{1}{2}\bigr)\frac{\pi}{5}\right)\]
is the boundary coupling $h(b)$ (\ref{eq:hbrelation}) in dimensionless
form, $G_{ij}=\langle i|j\rangle$ is the usual conformal metric on
the state space, and the matrix elements of the perturbing operators
\begin{equation}
B_{ij}=\langle i|\int_{0}^{\vartheta}d\vartheta\Phi\left(\mathrm{e}^{i\vartheta}\right)|j\rangle\quad,\quad(B_{R})_{ij}=\langle i|\varphi(1)|j\rangle\label{eq:Bmatrices}\end{equation}
are calculated on the upper half plane after the usual exponential
mapping from the strip as in \cite{DPTW1}.

\subsection{State identification and evaluation of matrix elements}

\begin{figure}
\begin{centering}
\psfrag{5}{$5$}\psfrag{10}{$10$}\psfrag{15}{$15$}\psfrag{20}{$20$}
\psfrag{2}{$2$}\psfrag{4}{$4$}\psfrag{6}{$6$}\psfrag{8}{$8$}
\psfrag{l}{$l$}\psfrag{e}{$e$}\includegraphics[width=14cm,keepaspectratio]{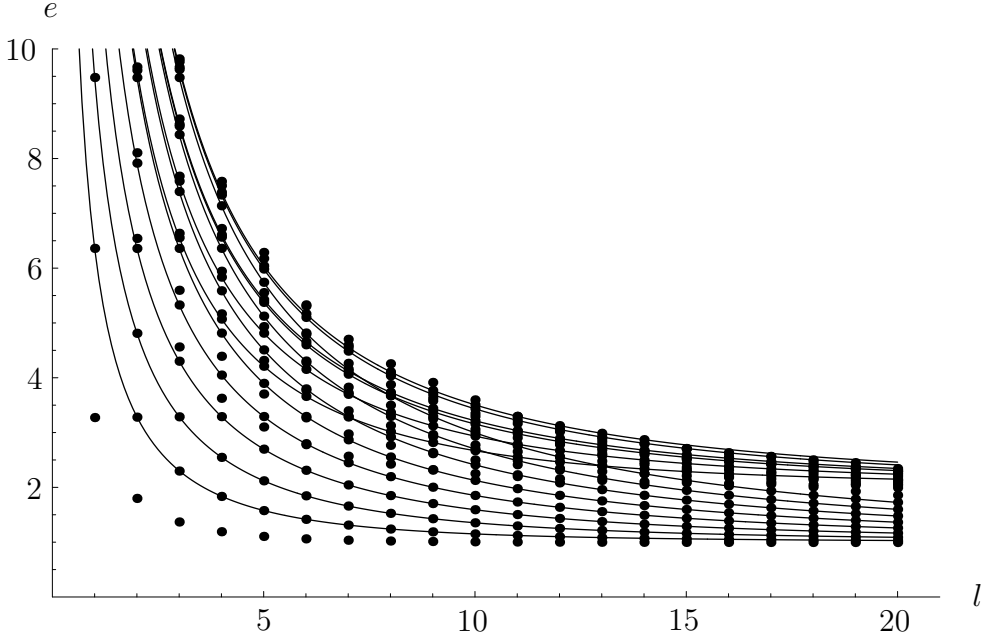}
\par\end{centering}

\caption{\label{fig:BTCSA-levels}BTCSA levels (at truncation level $25$)
and Bethe-Yang predictions for $b=-0.8$. Energy and volume are measured
in units of the bulk particle mass $m$ (i.e. $e=E/m$ and $l=mL$).
The discrete points are the BTCSA levels (with the ground state subtracted),
while the continuous lines are the Bethe-Yang predictions.}

\end{figure}

Once the BTCSA spectrum is obtained it is just a set of energy levels
represented as a bunch of numerical eigenvalues for a series of discreet
values of the volume parameter $l=mL$. We need to sort these levels
into lines parametrized by the volumes, labeled by their multi-particle
contents. This is best achieved by numerical comparison to levels
predicted by the Bethe-Yang equations (\ref{eq:bbye}), which is illustrated
in figure \ref{fig:BTCSA-levels}. The figure shows a very good agreement
between the two sets of data, however, the lowest state does not fit
any of the Bethe-Yang lines. The reason is that it corresponds to
a boundary excited state with energy\[
E_{*}=m\cos\frac{\pi(b+1)}{6}=0.99452\dots\times m\]
Some states corresponding to multi-particle states with the right
boundary in this excited state can be seen further up in the spectrum
as further points not fitting the Bethe-Yang lines. They can be described
if the reflection factor appearing in the Bethe-Yang equations (\ref{eq:bbye})
is replaced by the one valid for the excited state boundary, but we
omit the details here, since the agreement between the BTCSA spectrum
and the boundary scattering theory described in section 2.2 was already
thoroughly established in \cite{DPTW1}. 

Note that contrary to the case of the periodic TCSA where the Hilbert
space can be split into sectors according to total momentum, in BTCSA
there is a single sector. As a result, the dense part of the spectrum
(the {}``continuum'') already starts at the one-particle threshold,
which makes the identification of individual levels harder than it
was in the periodic case. We found a number of one-particle and two-particle
states, and could also identify the first two three-particle levels
over a volume range extending to around $l\sim14\dots17$, in marked
contrast with the periodic case where it was possible to perform a
systematic identification of levels with up to $4$ particles \cite{fftcsa1}
and even the lowest $5$-particle level was found \cite{fftcsa2}.
Even so we could generate a massive amount of useful data for each
value of the dimensionless parameter $b$.

Once we identified the levels, we extracted the corresponding eigenvectors
and evaluated the matrix elements of the boundary operator $\varphi$
using the method described in \cite{fftcsa1}. Suppose that we computed
two Hamiltonian eigenvectors as functions of the volume $L$:\begin{eqnarray*}
|\{I_{1},\dots,I_{n}\}\rangle_{L} & = & \sum_{i}\Psi_{i}(I_{1},\dots,I_{n};L)|i\rangle\\
|\{I_{1}',\dots,I_{k}'\}\rangle_{L} & = & \sum_{j}\Psi_{j}(I_{1}',\dots,I_{k}';L)|j\rangle\end{eqnarray*}
Let the inner products of these vectors with themselves be given by\begin{eqnarray*}
\mathcal{N} & = & \sum_{i,j}\Psi_{i}(I_{1},\dots,I_{n};L)G_{ij}\Psi_{j}(I_{1},\dots,I_{n};L)\\
\mathcal{N}' & = & \sum_{i,j}\Psi_{i}(I_{1}',\dots,I_{k}';L)G_{ij}\Psi_{j}(I_{1}',\dots,I_{k}';L)\end{eqnarray*}
where $G_{ij}=\langle i|j\rangle$ is the metric on the space of the
conformal state vectors $|i\rangle$ as before. Then the matrix elements
of the field $\varphi$ can be computed as

\begin{eqnarray}
 &  & m^{1/5}\langle\{I_{1}',\dots,I_{k}'\}\vert\varphi(0)\vert\{I_{1},\dots,I_{n}\}\rangle_{L}=\nonumber \\
 &  & \qquad\left(\frac{\pi}{l}\right)^{-1/5}\frac{1}{\sqrt{\mathcal{N}}}\frac{1}{\sqrt{\mathcal{N}'}}\sum_{j,l}\Psi_{j}(I_{1}',\dots,I_{k}';L)(B_{R})_{jl}\Psi_{l}(I_{1},\dots,I_{n};L)\label{eq:fftcsaevaluation}\end{eqnarray}
where $B_{R}$ is the boundary perturbation matrix entering the boundary
Hamiltonian (\ref{eq:btcsa_ham}), and the volume dependent prefactor
comes from the transformation of the primary field $\varphi$ under
the exponential map. The results will be reported in section 5, but
before that we turn to the estimation of the accuracy of the BTCSA
method.

\subsection{Numerical accuracy of BTCSA and error sources}

In order to understand what constitutes a satisfactory agreement between
the theoretical predictions and the numerical data, we need to understand
the possible sources of deviations. 

The truncation inherent in BTCSA introduces a specific source of error,
called the truncation error. It grows with the volume $L$ and also
it becomes larger when higher levels are considered. It is very hard
to control and estimate this error very precisely. It can be made
smaller by extrapolation procedures such as the one used in \cite{fftcsa1,fftcsa2}.
Recently a very efficient approach was introduced in the form of a
renormalization group under the variation of the truncation level
\cite{rgtcsa}. To avoid excessive numerical computations, we do not
use any extrapolation in the truncation level here; it turns out that
the highest truncation level we used ($25$, which gives a $434$
dimensional Hilbert space) is enough to achieve sufficient precision
provided we choose the parameter $b$ in a suitable range. We remark
that increasing the truncation level to the values used in the periodic
case \cite{fftcsa1,fftcsa2} (i.e. up to $30$) is not very practical
because the evaluation of the matrix elements of the bulk perturbation
is very time consuming and the required CPU time grows very fast. 

Another source of error comes from the fact that numerical diagonalization
becomes unstable where levels are nearly degenerate%
\footnote{Eigenvectors corresponding to degenerate or nearly degenerate levels
are very sensitive to any small perturbation, and thus even a small
truncation error can have a disproportionately large effect.%
}; an example of this phenomenon is shown in subsection 5.2.

A further reason for deviation is that every theoretical prediction
that we test against BTCSA is exact only up to residual finite size
corrections, i.e. contributions that decay exponentially with the
volume. As discussed in \cite{fftcsa1} for any measured quantity
there exists a volume range where the truncation errors and the residual
finite size corrections are of the same order; this is the so-called
scaling regime where the agreement between the theoretical predictions
and the numerical results is optimal. 

To see what precision can be expected in the scaling regime, we evaluated
three quantities characteristic of the ground state. The energy of
the ground state in finite volume has the large volume asymptotics\[
E_{0}(L)=\mathcal{B}m^{2}L+\mathcal{E}_{b}m+O\left(\mathrm{e}^{-\mu L}\right)\]
where the exact value of the bulk energy constant is \cite{lytba}\[
\mathcal{B}=-\frac{\sqrt{3}}{12}\]
and the boundary energy constant is given by \cite{DPTW1}\[
\mathcal{E}_{b}=\frac{\sqrt{3}-1}{2}+\sin\frac{\pi b}{6}\]
In addition, the finite volume vacuum expectation value has the asymptotics
\[
\left\langle 0|\varphi|0\right\rangle _{L}=\left\langle \varphi\right\rangle +O\left(\mathrm{e}^{-\mu L}\right)\]
where the exact asymptotic value $\left\langle \varphi\right\rangle $
is given by (\ref{eq:exactvev}).

\begin{table}
\begin{tabular}{|c|c||c|c||c|c|}
\hline 
$b$ & $\mathcal{B}$ (BTCSA) & $\mathcal{E}_{b}$ (exact) & $\mathcal{E}_{b}$ (BTCSA) & $m^{1/5}\left\langle \varphi\right\rangle $ (exact) & $m^{1/5}\left\langle \varphi\right\rangle $ (BTCSA)\tabularnewline
\hline
\hline 
$-2.3$ & $-0.144750$ & $-0.567555$ & $-0.568208$ & $-1.023503$ & $-1.043671$\tabularnewline
\hline 
$-2.0$ & $-0.144736$ & $-0.500000$ & $-0.499934$ & $-1.034417$ & $-1.052965$\tabularnewline
\hline 
$-1.7$ & $-0.144588$ & $-0.411121$ & $-0.411366$ & $-1.049802$ & $-1.066242$\tabularnewline
\hline 
$-1.4$ & $-0.144472$ & $-0.303105$ & $-0.303446$ & $-1.070303$ & $-1.083938$\tabularnewline
\hline 
$-1.1$ & $-0.144376$ & $-0.178614$ & $-0.178994$ & $-1.096875$ & $-1.106887$\tabularnewline
\hline 
$-1.0$ & $-0.144351$ & $-0.133975$ & $-0.134351$ & $-1.107309$ & $-1.115903$\tabularnewline
\hline 
$-0.8$ & $-0.144305$ & $-0.040711$ & $-0.041154$ & $-1.136265$ & $-1.130931$\tabularnewline
\hline 
$-0.5$ & $-0.144286$ & $0.107206$ & $0.106815$ & $-1.174595$ & $-1.174435$\tabularnewline
\hline 
$-0.2$ & $-0.144315$ & $0.261497$ & $0.261078$ & $-1.231176$ & $-1.226541$\tabularnewline
\hline 
$0.0$ & $-0.144372$ & $0.366025$ & $0.365596$ & $-1.278610$ & $-1.270723$\tabularnewline
\hline 
$0.3$ & $-0.144523$ & $0.522460$ & $0.522032$ & $-1.370592$ & $-1.356133$\tabularnewline
\hline
\end{tabular}

\caption{\label{tab:vactable} Boundary energy and vacuum expectation of $\varphi$:
exact predictions compared to BTCSA. The exact value of $\mathcal{B}$
is $-0.144338\dots$ .}

\end{table}

One can extract estimates for these quantities from the scaling regime
of BTCSA, which are compared with the predicted values in table \ref{tab:vactable}.
Besides noting the very good agreement it is useful to pay particular
attention to the vacuum expectation value, because it is in fact the
simplest (zero-particle) form factor that can be measured. In fact
we found that the deviation of the BTCSA determination of the matrix
elements from the theoretical predictions (\ref{eq:genffrelation},\ref{eq:diaggenrulesaleur})
follows well the precision of the determination of the vacuum expectation
value. We can therefore see that the highest precision can be achieved
when $-1.1\lesssim b\lesssim0.0$, and so it is in this range that
the numerical tests are the most reliable.

\section{Comparing theoretical predictions with BTCSA data}

In this section we present examples of the comparison between the
theoretical predictions (\ref{eq:genffrelation},\ref{eq:diaggenrulesaleur})
and BTCSA data. All the results presented below are for $b=-0.8$,
the same value of the model's parameter as in figure \ref{fig:BTCSA-levels},
which is in the optimum range for numerical precision. We actually
performed the calculations for fourteen different values of $b$ between
$-2.3$ and $+0.3$. We isolated the vacuum state, four one-particle,
four two-particle and two three-particle states for each value of
$b$, calculated matrix elements for all possible pairwise combinations
of these states, and found similarly good agreement as presented below.
The numerical deviations gradually increase when moving away from
the optimum range of $b$. In the optimum range, we found deviations
of the order $10^{-3}$ in the scaling regime; for the extremal values,
the deviations increased to a few percent for higher levels. There
was not a single matrix element in this huge set of data for which
any unexpectedly large deviation occurred; therefore we can state
that the numerical data are in full agreement with the theoretical
predictions.

We also recall (cf. \cite{fftcsa1}) that the relative phases of different
multi-particle states resulting from numerical diagonalization are
in general different from the conventions that follow from the form
factor axioms in subsection 2.1. Therefore the relations (\ref{eq:ffrelation},\ref{eq:genffrelation})
must be understood to hold up to some phase factors, and in the numerical
comparisons in figures \ref{fig:elementary-formfactors} and \ref{fig:Non-diagonal-matrix-elements}
we use the absolute values of the matrix elements. This issue does
not arise for the diagonal case as any such phase factor drops out
from (\ref{eq:diaggenrulesaleur}); in figure \ref{fig:Diagonal-matrix-elements}
we only removed a minus sign that appears on both side of this relation
after evaluation.

\subsection{Elementary form factors}

\begin{figure}
\noindent \begin{centering}
\psfrag{l}{$l$}\psfrag{ff}{$f$}
\psfrag{tcsa1}{$\langle 0|\varphi|\{1\}\rangle$}
\psfrag{tcsa2}{$\langle 0|\varphi|\{2\}\rangle$}
\psfrag{tcsa3}{$\langle 0|\varphi|\{3\}\rangle$}\subfigure[One-particle form factor]{\includegraphics[width=9cm,keepaspectratio]{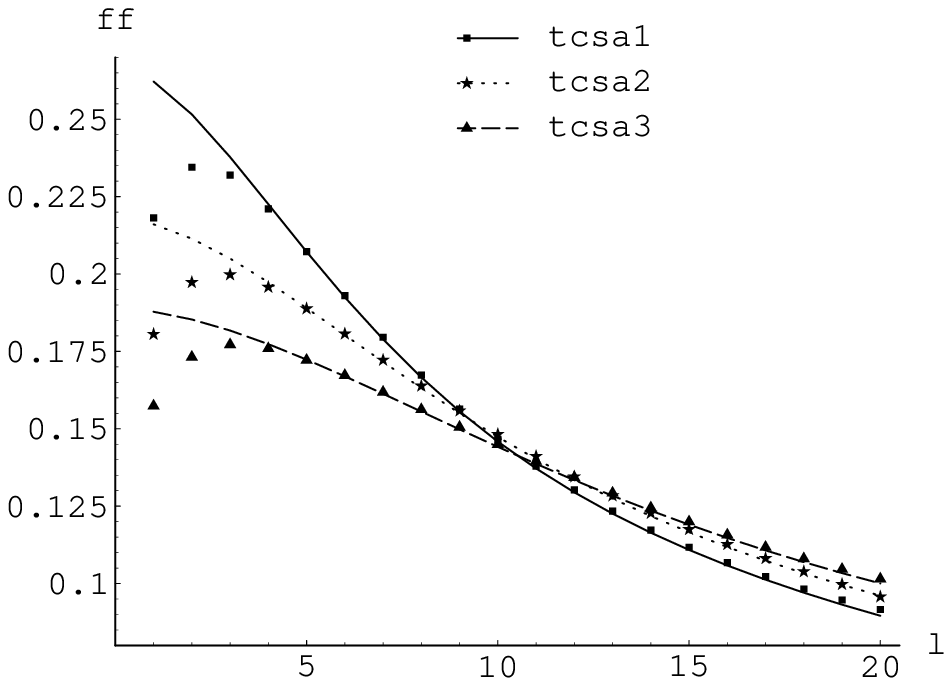}}\psfrag{l}{$l$}\psfrag{ff}{$f$}
\psfrag{tcsa1}{$\langle 0|\varphi|\{1,2\}\rangle$}
\psfrag{tcsa2}{$\langle 0|\varphi|\{1,3\}\rangle$}
\psfrag{tcsa3}{$\langle 0|\varphi|\{1,4\}\rangle$}\subfigure[Two-particle form factor]{\includegraphics[width=9cm,keepaspectratio]{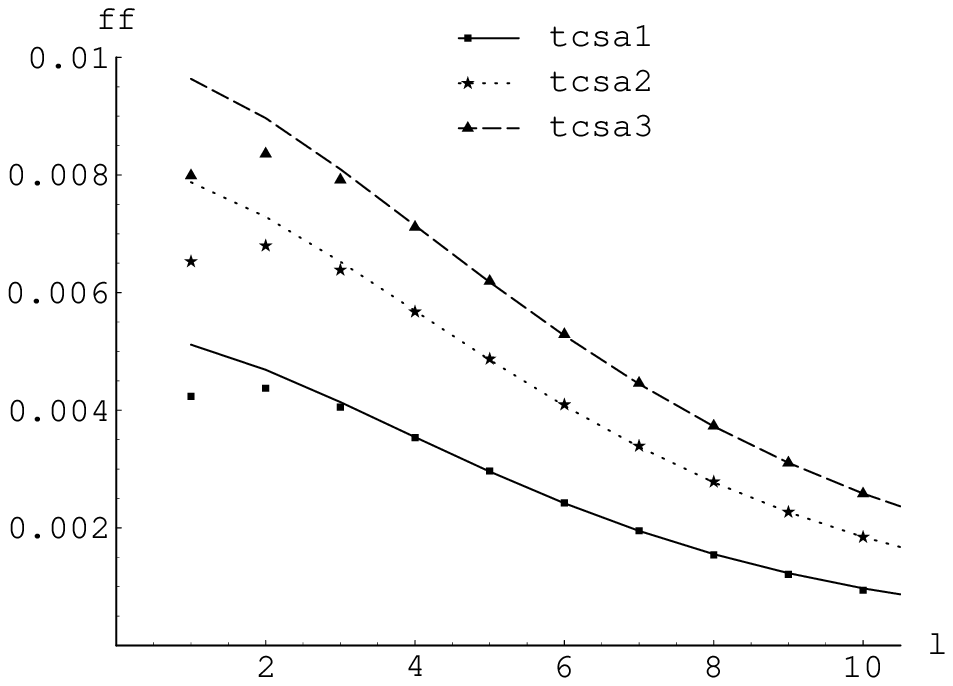}}
\par\end{centering}

\caption{\label{fig:elementary-formfactors}One-particle and two-particle elementary
form factors. $l=mL$ is the dimensionless volume parameter, while
$f$ is the magnitude of the matrix element in units of $m^{-1/5}$. }

\end{figure}

\begin{figure}
\begin{centering}
\psfrag{l}{$l$}\psfrag{ff}{$f$}
\psfrag{tcsa}{$\langle 0|\varphi|\{1\}\rangle$}
\psfrag{good}{interacting d.o.s.}
\psfrag{wrong}{naive d.o.s.}\includegraphics{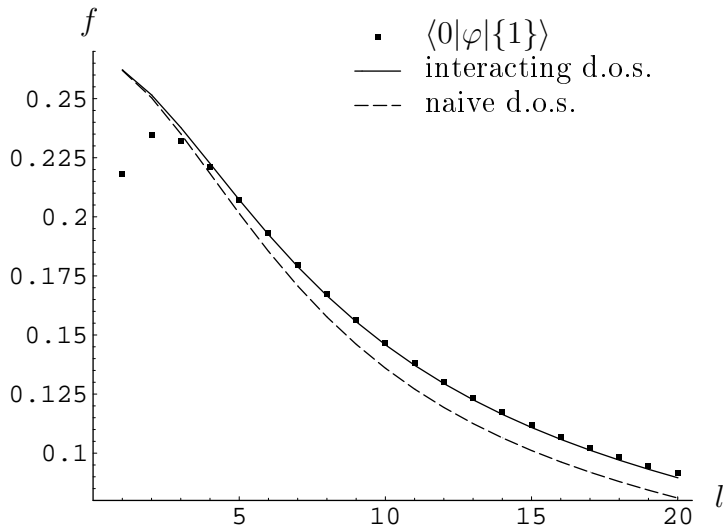}
\par\end{centering}

\caption{\label{fig:dos}The relevance of interaction corrections to the density
of states. The discrete points are TCSA data for the one-particle
matrix element shown, continous line is the prediction (\ref{eq:ffrelation}),
while the dashed line shows the same with the naive density of states.}

\end{figure}

Elementary form factors are defined in eqn. (\ref{eq:elementaryff});
these are the ones which enter the form factor equations given in
subsection 2.1, and all other form factors can be obtained from them
by crossing. Comparing such matrix elements against BTCSA provide
a test for form factor functions with all their arguments real, according
to eqn. (\ref{eq:ffrelation}). However, the numerical magnitude of
the corresponding finite volume matrix elements decreases very fast
with the number of particles. The three-particle matrix elements are
already too small to be measured meaningfully from BTCSA (they are
of the order of truncation errors), and therefore we limited ourselves
to the one- and two-particle cases, shown in figure \ref{fig:elementary-formfactors}.
Note that the two-particle matrix elements are already smaller than
the one-particle ones by more than an order of magnitude; accordingly,
the relative precision in their case is around one percent, while
for the one-particle matrix elements we could achieve around $10^{-3}$
(or even slightly better) in the scaling regime. According to the
discussion at the end of subsection 3.2 this is enough to test that
the inclusion of the phase-shifts in (\ref{eq:dosfinvol}) which describe
the interaction between the particles and also with the boundary is
necessary to achieve agreement between the theoretical predictions
(\ref{eq:ffrelation}) and truncated conformal space data. This is
illustrated in figure \ref{fig:dos} where besides plotting the correct
prediction (\ref{eq:ffrelation}) involving the full one-particle
density of states \[
\rho_{1}(\theta)=2mL\cosh\theta+\frac{\partial\delta^{(\alpha)}(\theta)}{\partial\theta}+\frac{\partial\delta^{(\beta)}(\theta)}{\partial\theta}\]
 which takes into account interactions as defined in (\ref{eq:dosfinvol}),
we also plot a version where the density of states is substituted
by\[
\rho_{1}^{\mathrm{naive}}(\theta)=2mL\cosh\theta\]
Here $\delta^{(\alpha)}$ and $\delta^{(\beta)}$ are the boundary
phase-shifts define in (\ref{eq:bphaseshifts}). Note that for small
$L$ the difference between the naive and the full phase-shift goes
to $0$. This may seem surprising at first, but it is due to the fact
that $\theta$ increases with decreasing $L$ and the phase-shift
derivatives decrease exponentally for $\theta$ away from $0$. Although
figure \ref{fig:dos} only shows the case of the first one-particle
state, similar results are obtained for all the form factors discussed
in this paper.

\subsection{Non-diagonal matrix elements}

\begin{figure}
\noindent \begin{centering}
\psfrag{l}{$l$}\psfrag{ff}{$f$}
\psfrag{tcsa1}{$\langle \{1\}|\varphi|\{1,2\}\rangle$}
\psfrag{tcsa2}{$\langle \{1\}|\varphi|\{1,3\}\rangle$}
\psfrag{tcsa3}{$\langle \{2\}|\varphi|\{1,2\}\rangle$}
\psfrag{tcsa4}{$\langle \{2\}|\varphi|\{1,3\}\rangle$}\subfigure[1-particle---2-particle]{\includegraphics[width=9cm,keepaspectratio]{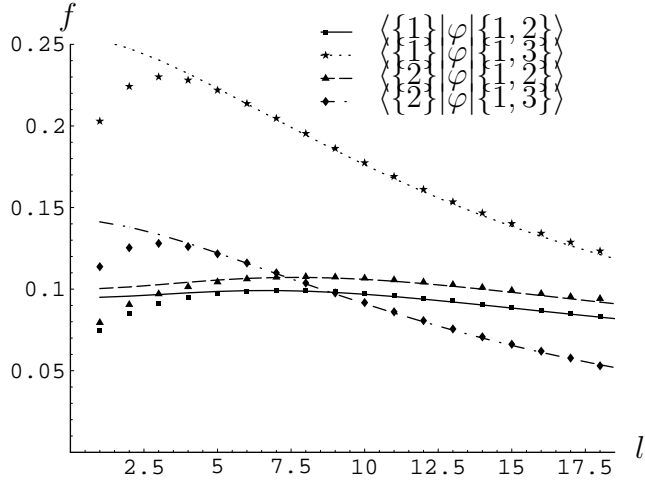}}\psfrag{l}{$l$}\psfrag{ff}{$f$}
\psfrag{tcsa1}{$\langle \{1,2\}|\varphi|\{1,3\}\rangle$}
\psfrag{tcsa2}{$\langle \{1,2\}|\varphi|\{1,4\}\rangle$}
\psfrag{tcsa3}{$\langle \{1,3\}|\varphi|\{1,4\}\rangle$}
\psfrag{tcsa4}{$\langle \{1,3\}|\varphi|\{2,3\}\rangle$}\subfigure[2-particle--2-particle]{\includegraphics[width=9cm,keepaspectratio]{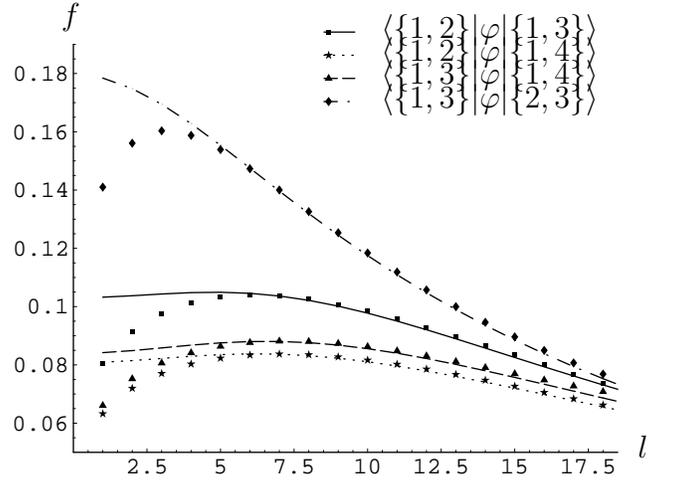}}\\
\psfrag{l}{$l$}\psfrag{ff}{$f$}
\psfrag{tcsa1}{$\langle \{1,2\}|\varphi|\{1,2,3\}\rangle$}
\psfrag{tcsa2}{$\langle \{1,3\}|\varphi|\{1,2,3\}\rangle$}
\psfrag{tcsa3}{$\langle \{2,3\}|\varphi|\{1,2,3\}\rangle$}
\psfrag{tcsa4}{$\langle \{2,3\}|\varphi|\{1,2,4\}\rangle$}\subfigure[2-particle--3-particle]{\includegraphics[width=9cm,keepaspectratio]{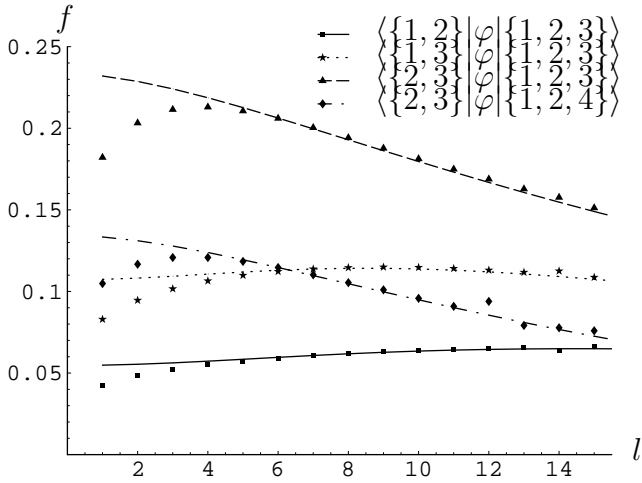}}\psfrag{l}{$l$}\psfrag{ff}{$f$}
\psfrag{tcsa1}{$\langle \{1,2,3\}|\varphi|\{1,2,4\}\rangle$}\subfigure[3-particle--3-particle]{\includegraphics[width=9cm,keepaspectratio]{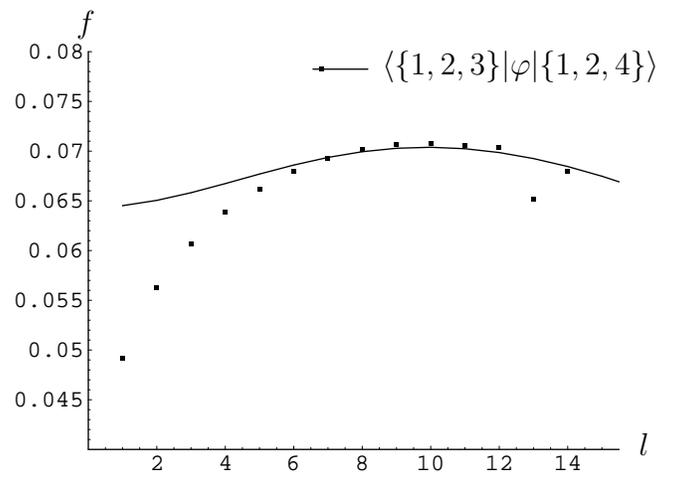}}
\par\end{centering}

\caption{\label{fig:Non-diagonal-matrix-elements}Non-diagonal matrix elements.
$l=mL$ is the dimensionless volume parameter, while $f$ is the magnitude
of the matrix element in units of $m^{-1/5}$. }

\end{figure}

Form factor functions with more than two particles can only be measured
with sufficient accuracy from more general (non-diagonal) matrix elements,
for which the theoretical prediction is given by eqn. (\ref{eq:genffrelation}).
We present such data in figure \ref{fig:Non-diagonal-matrix-elements},
which provide a check for form factor functions involving up to 6
particles; the relative deviation between the BTCSA data and the theoretical
prediction (\ref{eq:genffrelation}) is around $10^{-3}$ in the scaling
regime. We omitted the case of 1-particle--1-particle matrix elements,
because they test the two-particle form factor already checked above;
they show similarly good agreement. We also omitted the case of 1-particle--3-particle
matrix elements as they are again too small to be extracted with high
enough precision. 

We remark that there is an easily visible, quite large deviation in
figure \ref{fig:Non-diagonal-matrix-elements} (c) at $l=13$. The
reason is that the corresponding matrix element involves the second
three-particle level $|\{1,2,4\}\rangle$, which crosses another level
in the vicinity. As we already discussed at the beginning of subsection
4.2, in the vicinity of such level crossings the pair of levels can
bifurcate into a pair of complex conjugate eigenvalues, which increases
the numerical deviation due to truncation errors.

\begin{figure}
\noindent \begin{centering}
\psfrag{ee}{$\Re e\,e$}
\psfrag{l}{$l$}\includegraphics[width=7cm]{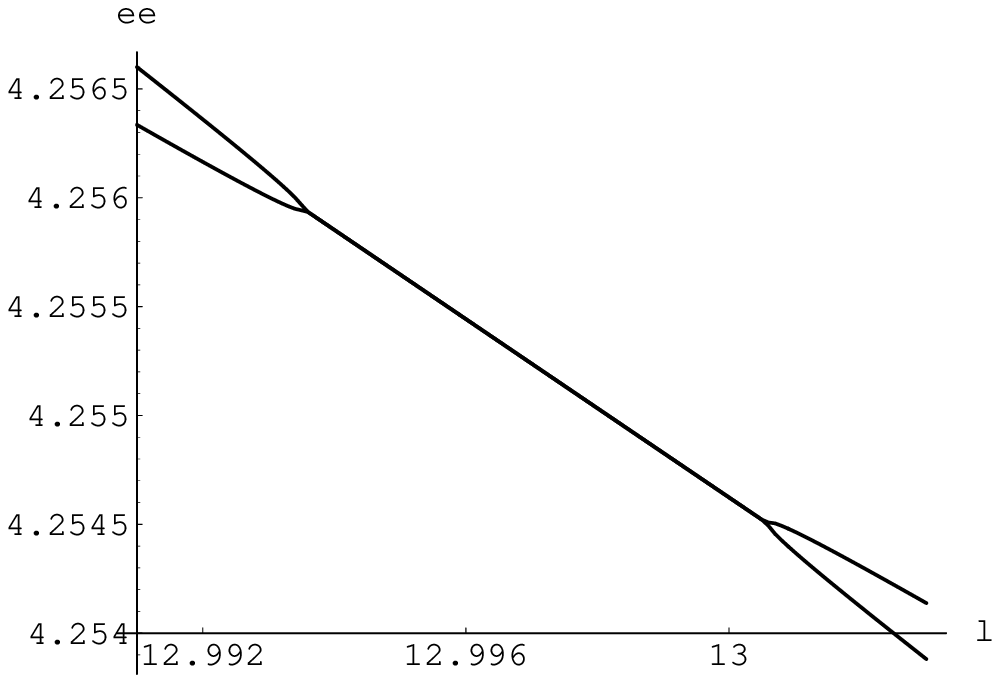}\psfrag{ee}{$\Im m\,e$}
\psfrag{l}{$l$}\includegraphics[width=7cm]{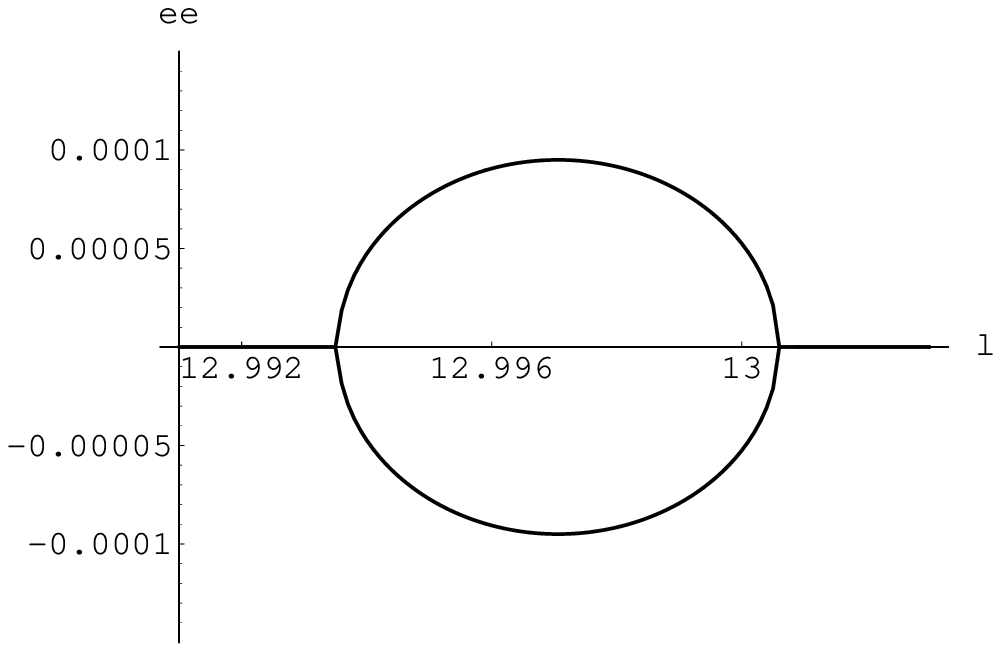}
\par\end{centering}

\caption{\label{fig:Level-crossing-around} Level crossing around $l=13$.
Energy ($e$) is measured in units of $m$, relative to the ground
state. The two figures show the real and the imaginary parts of the
two levels, respectively.}

\end{figure}

This particular level crossing is shown in figure \ref{fig:Level-crossing-around},
and it can be easily seen that the bifurcation is indeed due to truncation
errors. At such a large volume the Bethe-Yang equations (\ref{eq:bbye})
give a very accurate description of energy levels, since the residual
finite size corrections are very small, of order $\mbox{e}^{-l}\sim10^{-6}$.
Therefore imaginary parts of levels which are larger than this order
of magnitude can only be due to truncation errors. Indeed the imaginary
parts at the middle of the level crossing are of order $10^{-4}$
which is also consistent with the magnitude of truncation errors that
can be estimated from the deviation between the Bethe-Yang and BTCSA
levels around this point.

Turning to an exact description of the finite size spectrum, for periodic
boundary conditions it can be proved that the full exact finite volume
spectrum of the scaling Lee-Yang model is strictly real \cite{rsos},
and therefore the similar lacunae observed by Yurov and Zamolodchikov
\cite{yurov_zamolodchikov} can only be due to truncation. We expect
that a similar result holds for the scaling Lee-Yang model with boundaries
when $b$ is real i.e. when $\left|h\right|<h_{crit}$%
\footnote{For larger values of the boundary coupling $h$ the ground state is
destabilized and the spectrum turns complex \cite{DPTW1}.%
}, and a proof can probably be given based on the boundary TBA description
of the finite volume spectrum established in \cite{DPTW1}, but we
do not pursue this issue further here.

\subsection{Diagonal matrix elements}

\begin{figure}
\noindent \begin{centering}
\psfrag{l}{$l$}\psfrag{ff}{$f$}
\psfrag{tcsa1}{$\langle \{1\}|\varphi|\{1\}\rangle$}
\psfrag{tcsa2}{$\langle \{2\}|\varphi|\{2\}\rangle$}
\psfrag{tcsa3}{$\langle \{3\}|\varphi|\{3\}\rangle$}\subfigure[1-particle]{\includegraphics[width=9cm,keepaspectratio]{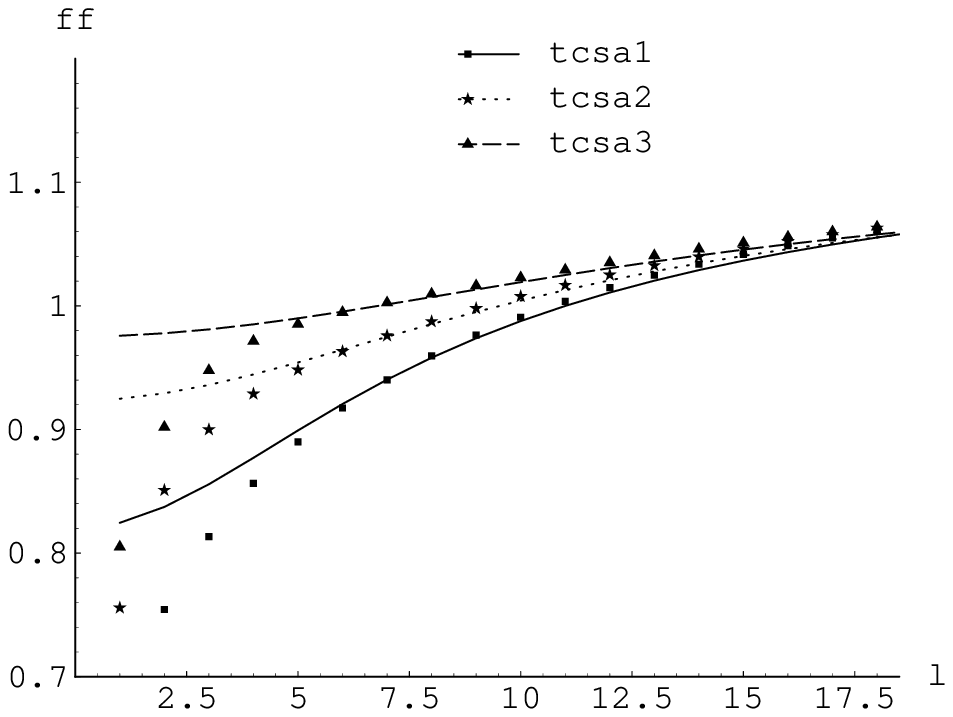}}\psfrag{l}{$l$}\psfrag{ff}{$f$}
\psfrag{tcsa1}{$\langle \{1,2\}|\varphi|\{1,2\}\rangle$}
\psfrag{tcsa2}{$\langle \{1,3\}|\varphi|\{1,3\}\rangle$}
\psfrag{tcsa3}{$\langle \{1,4\}|\varphi|\{1,4\}\rangle$}\subfigure[2-particle]{\includegraphics[width=9cm,keepaspectratio]{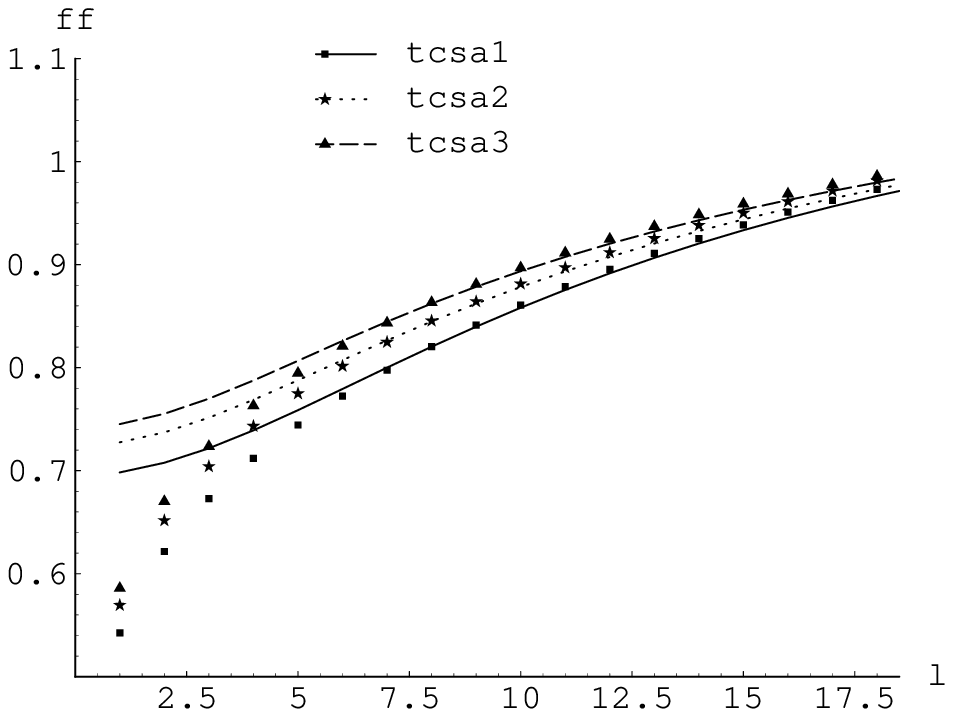}}\\
\psfrag{l}{$l$}\psfrag{ff}{$f$}
\psfrag{tcsa1}{$\langle \{1,2,3\}|\varphi|\{1,2,3\}\rangle$}
\psfrag{tcsa2}{$\langle \{1,2,4\}|\varphi|\{1,2,4\}\rangle$}\subfigure[3-particle]{\includegraphics[width=9cm,keepaspectratio]{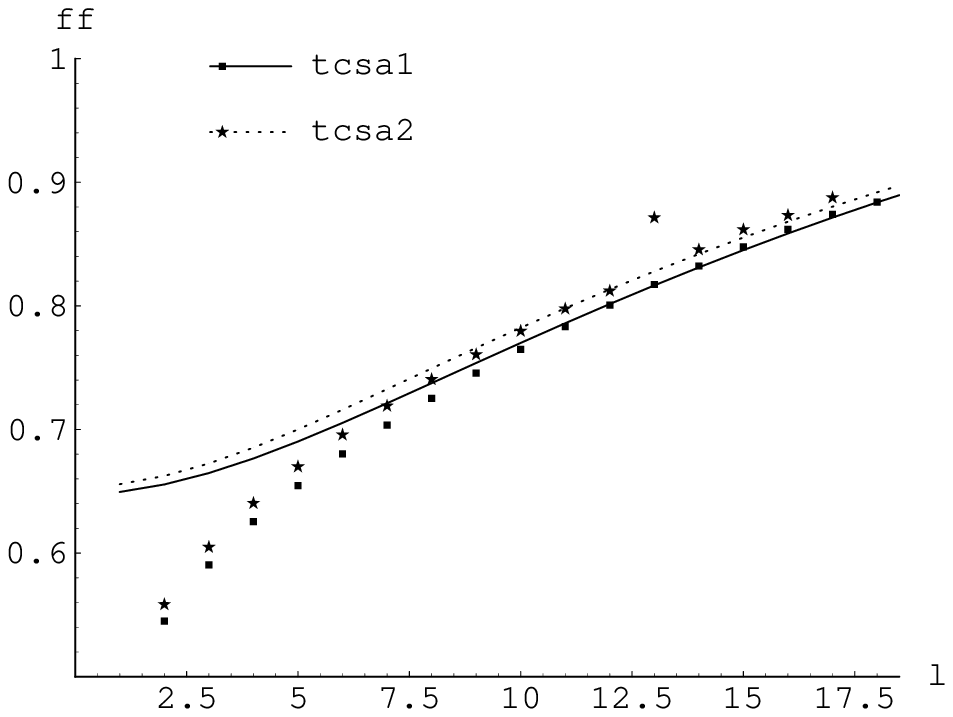}}
\par\end{centering}

\caption{\label{fig:Diagonal-matrix-elements}Diagonal matrix elements. $l=mL$
is the dimensionless volume parameter, while $f$ is the matrix element
in units of $m^{-1/5}$ , multiplied by $-1$ (both in the case of
the BTCSA data and the theoretical prediction).}

\end{figure}

The comparisons in the previous two subsections can be considered
as a direct verification of the form factor functions given in subsection
2.3. Once we are certain that the elementary form factor functions
are correct, we can view the comparison of diagonal matrix elements
as testing the structure of the disconnected terms involved in (\ref{eq:diaggenrulesaleur}).
The data are presented in figure \ref{fig:Diagonal-matrix-elements}
and again show excellent agreement (with deviations of order $10^{-3}$
in the scaling regime) apart from the presence of the deviation due
to the level crossing discussed in the previous subsection.

\section{Conclusions}

In this paper we have succeeded to extend the description of finite
volume matrix elements of local operators \cite{fftcsa1,fftcsa2}
to the case of boundary operators in a boundary quantum field theory.
We have compared the predictions of this framework to data extracted
from the boundary truncated conformal space approach, and found excellent
agreement. This provides a very strong evidence both for the validity
of the boundary form factor bootstrap introduced in \cite{bffprogram},
and to the description of finite volume matrix elements in terms of
the infinite volume form factors given by eqns. (\ref{eq:genffrelation},\ref{eq:diaggenrulesaleur}).

As pointed out in \cite{fftcsa2}, understanding finite size corrections
to form factors is not only valuable as a tool to check the form factor
bootstrap, but also for the calculation of finite temperature correlators.
The introduction of finite volume regularizes the singular terms that
come from disconnected pieces, and makes it possible to develop a
systematic low-energy expansion for finite temperature correlators,
as demonstrated on the example of one-point functions in \cite{fftcsa2}.
A form factor expansion for the finite temperature expectation values
of boundary operators has already been developed in \cite{ftop};
we also plan to extend these results to two-point correlation functions.
It was already demonstrated in the seminal paper by Affleck and Ludwig
\cite{affleckludwig} that the finite temperature setting is essential
in studying boundary renormalization group flows, and therefore we
expect that such an expansion can be useful in this context.

\subsection*{Acknowledgments}

GT wishes to thank Z. Bajnok and L. Palla for useful discussions.
This research was partially supported by the Hungarian research fund
OTKA K60040. GT was also supported by a Bolyai J\'anos research scholarship.

\end{document}